\documentclass[draft]{agujournal2019}
\usepackage{aas_macros}
\usepackage{url} 
\usepackage{lineno}
\usepackage{xcolor}
\usepackage[table]{xcolor}
\usepackage[inline]{trackchanges} 

\newif\ifshowchanges
\showchangesfalse

\ifshowchanges
  \newcommand\hll[1]{\bgroup\hskip0pt\color{red!80!black}#1\egroup}
\else
  \newcommand\hll[1]{#1}
\fi

\draftfalse

\journalname{JGR Space Physics}

\begin{document}

\title{Atmospheric Mass Flux as a Function of Ionospheric Emission on Unmagnetized Earth}

\authors{P.C. Hinton \affil{1,2,3}, D.A. Brain \affil{1,2}, N.R. Schnepf  \affil{4}, R. Jarvinen \affil{5}, J. Cessna \affil{2}, F. Bagenal \affil{1,2}}

\affiliation{1}{University of Colorado, Boulder Colorado, USA}
\affiliation{2}{Laboratory for Atmosphere and Space Physics, Boulder Colorado, USA}
\affiliation{3}{National Solar Observatory, Boulder Colorado, USA}
\affiliation{4}{Geologic Hazards Science Center, United States Geologic Survey, Golden, CO}
\affiliation{5}{Finnish Meteorological Institute, Helsinki, Finland}

\correspondingauthor{Parker Hinton}{parker.hinton@colorado.edu} \begin{keypoints}
\item For constant solar driving conditions, the mass content of unmagnetized Earth's atmosphere \hll{may} be stable within 3\% over a billion years.
\item Increased ionospheric emission inflates the magnetosphere via mass-loading and more efficiently deflects the solar wind. 
\item It may be possible for planetary upper atmospheres to accrete more mass from stellar winds than what escapes from the atmosphere to space.
\end{keypoints}

\begin{abstract}
We explore ion escape from, and solar ion deposition to, \hll{an unmagnetized Earth-like planet}. We use RHybrid, an ion-kinetic electron-fluid code to simulate the global plasma interaction of unmagnetized Earth with the solar wind. We vary the global ionospheric emission rate, and quantify the resultant planetary ion escape rates ($O^+$ and $H^+$) and the solar wind deposition rate ($H^+$). We use these results to compute the net mass flux to the atmosphere and find that the solar ion deposition rate could be comparable to planetary ion escape rates. For the emission rates simulated, our results show that under typical solar wind conditions ($v_{sw} = 400 \ km \ s^{-1}$, $n_{sw} = 5 \ cm^{-3}$), the mass of the atmosphere would decrease by less than 3\% over a billion years, indicating that Earth's intrinsic magnetic field may be unnecessary for retention of its atmosphere. Lastly, we present a hypothesis suggesting that ionospheric emission may evolve through time towards a critical emission rate that occurs at a net mass flux of zero.
\end{abstract}
%
%
\section{Plain Language Summary}

The electromagnetic interaction between a planet's ionosphere and the solar wind can result in a planet losing atmospheric mass to space. We simulate this interaction for an Earth-like planet in the case that the planet has no magnetic field. We detail how the plasma interaction changes as the ionospheric emission rate is changed. We quantify the rate at which ions escape the planet, as well as the rate that solar wind ions impact the atmosphere. We find that, for the various ionospheric emission rates tested, the planet would lose less than $3\%$ of an Earth's atmosphere worth of mass over a billion years. These findings indicate that a magnetic field is not required in order for atmospheric retention.

\section{Introduction}

Planetary atmospheres can escape to space over time, which can alter the planet's habitability and surface morphology. Geologic evidence indicates that Mars' surface previously hosted oceans and other liquid water features \cite{Carr1992, Carr2003}. \hll{For water to have been stable on the Martian surface would have required an atmospheric pressure of at least $0.5 - 1 \ bar$, which is much more dense than the $6.1 \ mbar$ observed today \cite{Carr1999, Kite2021}.} The decrease in atmospheric pressure suggests that most of Mars' atmosphere has escaped to space or has been absorbed by the surface. Measurements of the $^{38}Ar/^{36}Ar$ isotope ratio in the Martian upper atmosphere have been used to estimate that as much as $66 \pm 5 \%$ of atmospheric argon has been lost to space over the last $\sim 4$ billion years \cite{Jakosky2017}. \hll{However, modeled estimates of past escape for any given period of time are sensitive to uncertain input parameters such as solar wind properties \cite{Lammer2013, Jakosky2018}. Given this uncertainty, one common starting point for understanding integrated losses is to assume a constant escape rate across time \cite{Persson2020}.}

Atmospheric escape processes can broadly be broken down into escape of neutrals and escape of ions \cite{Brain2016}. Neutral escape can be further categorized as thermal (Jeans) and non-thermal escape \cite{Chaufray2024}. \hll{Refer to \citeA{Gronoff2020} for a thorough overview of different escape mechanisms.} The dominant escape mechanism varies based on the particular species and planet in question. For example, the primary neutral escape mechanism for oxygen on Mars today is dissociative recombination of molecular ions ($e^- + O^+_2 \rightarrow O + O$) \cite{Jakosky2018, Lillis2017}. Due to Mars' lower gravity, this chemical process results in fast neutral oxygen atoms with $v > v_{esc}$. However, this mechanism is not a dominant driver of escape on Venus or Earth due to the stronger gravitational fields \cite{Brain2016, Dubinin2011}.

Ion escape may sometimes be the dominant escape pathway for certain species on more massive planets \cite{Dubinin2011}. On Venus, oxygen ion escape is measured to be $(3-10) \times10^{24} \ s^{-1}$ \cite{Futaana2017, Ramstad2021}, whereas neutral escape of oxygen is estimated to be about $25 \%$ of this rate \cite{Lammer2006}. On Earth, oxygen escapes primarily as an ion via the polar wind \cite{Ramstad2021, Yau1997} and neutral escape of hydrogen is governed by Jeans escape \cite{Clarke2024, Gronoff2020}. In the current study, we treat only ion escape and deposition in our modeling of an Earth-like planet.

A planet with a sufficiently powerful internal dynamo, such as present-day Earth, will have an intrinsic magnetic field reaching multiple planetary radii into space \cite{Elsasser1950}. It is often believed that such intrinsic magnetospheres protect the planet's atmosphere from solar wind erosion \cite{Bertrand2021, Peccerillo2021, Bennet2020}. The existence of an intrinsic magnetic field on Earth and the lack of a significant intrinsic field on Mars has been used to explain why Earth hosts a dense atmosphere while Mars does not \cite{Jakosky2018, Jakosky2022}. However, the solar wind interaction with a planetary ionosphere can induce a magnetosphere around unmagnetized terrestrial planets, which may provide a similar level of atmospheric shielding afforded by an intrinsic magnetic field \cite{Johnson1969, Gunell2018}. \citeA{Birk2004} found that unmagnetized Earth will generate a comparatively strong magnetic field via induction in the ionosphere from the solar wind interaction \hll{with a peak intensity of $\sim 0.3 \ Ga$}.

With such a small sample size of planets available for direct observation, it is difficult to resolve the role of an intrinsic magnetic field with in-situ spacecraft measurements of atmospheric escape \cite{Ramstad2021}. Hybrid modeling \cite{Kallio2012,Egan2019}, magnetohydrodynamic (MHD) modeling \cite{Sakata2020}, and analytical approaches \cite{Gunell2018} that quantify escape as a function of intrinsic magnetic field strength find that in some regimes a planetary magnetic field can be protective, while in others it may actually enhance global escape. In the present study, we reduce Earth's magnetic dipole moment to $\vec m_{dp} = 0$. This represents the case where the geodynamo is shut off. In such a scenario, accretion of the solar wind may become a relevant variable.

The idea of solar wind particles providing a mass source term for the atmosphere of a planet is nearly as old as the discovery of the solar wind itself \cite{DeTurville1961}. It was theorized that the solar wind would probably not constitute a significant source term for planets with strong intrinsic magnetic fields, however, its relevance for unmagnetized bodies such as Mars and Venus was left as an open question \cite{Johnson1969}. Theoretical work determining the different regimes in which solar wind deposition may be important is lacking, partially due to the complexity of the interaction.

Mars Express (\emph{MEX}) observed charged particles consistent with a solar wind origin all the way down to the spacecraft's periapsis of $250 \ km$ \cite{Lundin2009}. The Mars Atmosphere and Volatile Evolution (\emph{MAVEN}) spacecraft also sees, on every single orbit through the dayside Martian upper atmosphere, a population of protons with solar wind velocity \cite{Halekas2015}. Were the Earth to become unmagnetized, the solar wind bow shock would occur much closer to the planet than it currently does, allowing for a much more direct interaction between the solar wind and the planet. Unmagnetized Earth, with a larger cross-section and higher solar dynamic pressure than Mars, would also reasonably experience solar wind deposition into its upper atmosphere. Therefore, including solar wind accretion may be important when trying to understand the role of an intrinsic magnetic field with respect to atmospheric flux.  

Global plasma hybrid codes self-consistently describe the formation of an induced magnetosphere and resolve kinetic trajectories of both solar and planetary ions \cite{Ledvina2008, Brain2010, Egan2018}. We use RHybrid \cite{Jarvinen2018}, an ion-kinetic particle-in-cell code, to simulate unmagnetized Earth's global plasma interaction with the solar wind. Instead of solar ionizing flux, the global ionospheric outflow rate is a free parameter in the model that prescribes ion emission from an inner boundary above the ionosphere. The RHybrid model and its predecessor HYB has been used extensively at Venus \cite{Jarvinen2013, Jarvinen2020}, Mars \cite{Jarvinen2018, Egan2018}, and for unmagnetized terrestrial planets with varying radii \cite{Hinton2024}.

A global ion outflow rate is the number of ions per second that cross a specified altitude in the anti-planet direction. At the unmagnetized planets, outflowing ions near the top of the ionosphere are subjected to electromagnetic interaction with the shocked solar wind that drives escaping ion populations. However, particle-tracing on the day-side of Mars indicates that most outflowing photo-ions end up recirculating back to the ionosphere \cite{Fang2008, Fang2010}. Thus, it can be important to distinguish between outflowing ions and ions that go on to escape the planet, however, outflows measured sufficiently far from the planet can be considered to be escaping. We define the near-planet outflow rate for species $m$ at the `top' of the ionosphere to be the ionospheric emission rate ($E_m$).

In this study, we seek to understand how atmospheric ion escape and solar wind deposition vary as a function of the emission rate for the case of unmagnetized Earth. We run six simulations varying only $E_m$. Each simulation includes planetary $H^+$ and $O^+$, and solar $H^+$. We quantify the inflation of the induced magnetosphere and the resultant planetary escape and solar wind deposition. From these results, we compute the net mass flux of the atmosphere and explore the significance of ion escape on geologic timescales. Lastly, we present a hypothesis suggesting that, all else being equal, the ionospheric emission rate may evolve towards a critical value where the net mass flux to the upper atmosphere is zero. The operative mechanism is that as net mass is added to the upper atmosphere it may begin to inflate. This can increase total ion production, which will increase the ion loss rates. Whereas, as net mass is lost, the opposite may occur. This suggests that variations from a net mass flux of zero can have negative feedback.

\section{Methods}
\subsection{RHybrid: General Description}

RHybrid is a global model for planetary plasma interactions that treats ions as macroscopic ion clouds, which are representative of many real physical ions, and electrons as a charge neutralizing fluid. Charge exchange, electron impact ionization, and ionospheric photochemistry are not directly implemented in RHybrid. RHybrid uses a planet-centered Cartesian coordinate system and the simulation domain is a rectangular cuboid where electric and magnetic fields are stored on cube-shaped grid cells.

The initial condition of the model is particle free. The upstream undisturbed solar wind is prescribed and injected in the simulation on the far +x side of the domain with a frozen-in perpendicular IMF component to the flow. Geocentric Solar Ecliptic (GSE) coordinates are used. The x-axis is the planet-sun line. The y-axis lies in the ecliptic plane pointing in the anti-orbital direction, i.e. the y-axis points in the clockwise direction if viewed `from above'. In this study, this is also the direction of the convection electric field. The z-axis is normal to the ecliptic and points along the rotation axis assuming the planet has an obliquity of $0 ^{\circ}$. The solar wind propagates towards the planetary ionospheric obstacle, which is modeled as a superconducting sphere with a resistivity of zero and an upwards emission of ionospheric ion species. The magnetic field is advanced with Faraday's law [$\frac{\partial \vec B}{\partial t} =- \vec \nabla \times \vec E$] and the electric field is determined by Ohm's law [$\vec E = -\vec U_e\times \vec B + \eta _a \vec J + \vec E_p$, where $J$ is the current density, $U_e$ is the electron bulk velocity, $\eta_a$ is the explicit resistivity, and $E_p$ is the electric field due to electron pressure]. The ions are propagated self-consistently via the Lorentz force [$m_i\frac{d \vec v_i}{dt}=q_i(\vec E + \vec v_i \times \vec B)$]. We provide a brief description here so that this work is self-contained; further descriptions of the RHybrid code can also be found in  recent publications \cite{Jarvinen2018,Jarvinen2020,Jarvinen2022}. The code is open source and can be found at github.com/fmihpc/rhybrid.

In a planetary ionosphere, the dominant form of ion production is photoionization and subsequent electron impact ionization \cite{Haider2011, Richards2011}. In an atmosphere, the production rate $p$ of species $m$, considering only photoionization, at altitude $z$ is defined as:

\begin{linenomath*}
\begin{equation}
p_{m}(z) =  \int_{0}^{\lambda_M} F_\lambda (z) n_{n,m}(z)  \sigma_\lambda  \,d\lambda \hspace{10mm} [ions \ cm^{-3}s^{-1}]
\end{equation}
\end{linenomath*}

where $n_{n,m}(z)$ is the number density of neutrals, $\lambda$ is wavelength, $\lambda_M$ is the ionization threshold, $F$ is the photon flux, and $\sigma$ is the photo-absorption cross-section (\citeA{Cravens1997}, Section 7.3.3). Once created, the ion can diffuse upward, downward, or laterally. RHybrid does not include production mechanisms for ionospheric ions, but instead tracks the trajectories of ions once they're created. Upward traveling ions constitute the ionospheric emission rate; this population of particles is the primary input to RHybrid.

At a prescribed radius from the center of the planet, called the emission radius, RHybrid injects an outward flux of planetary ions from a spherical shell around the planet. The global emission rate $E_m$ is prescribed for each species $m$, and the emission at any location is a function of solar-zenith angle (SZA) $\theta$ \cite{Schunk2009}. On the day side of the planet, ion emission is modulated by an SZA-dependent scale factor $a$: $a(\theta) = 1 - 0.9(1-cos(\theta))$. This ensures maximum emission at the subsolar point. On the nightside of the planet, (terminator at $\theta = 90^\circ$), $a$ is defined to be $0.1$.

\subsection{Simulation Parameters}

\hll{We inject solar wind macro-ions on the $yz$ plane at the far $x$ side of the domain with velocity $v = -400 \ km/s \ \hat{x}$, magnetic field $B = -5 \ nT \ \hat{z}$, and proton density $n = 5 \ cm^{-3}$}. The choice of magnetic field direction conveniently aligns the y-axis with the convection electric field and puts the solar wind magnetic field in the `top-down' configuration. We use spatial resolution $dx = 411km$, temporal resolution $dt = 0.026s$, and a domain that is a Cartesian cube with a side length of $l = 16 \ R_E$. The planet is centered at $x=0$, $y=0$, $z=0$, but the domain is $x = [-12,4]R_E$, providing three times the simulation space in the downwind direction as opposed to the upwind direction (see Figure 1 for visualization). We run each simulation on 256 cores (4 nodes, 64 cores per node) for 7 days, which is the wall clock time of the University of Colorado's Alpine supercomputer, totaling 41,328 cpu hours per run. For context, a standard 4-core computer would take over a year to complete a single one of the 6 runs in this study.

The planet is given mass $m = 5.97\times 10^{24} \ kg$ ($m_E$) and radius $R_p = 6,371 \ km$ ($R_E$). The force of gravity from the planet is included in the simulation runs. Explicit resistivity, allowing the diffusion of the magnetic field above $R_E + 200 \ km$, is $\eta_a=\eta_c\mu_0 dx^2/dt$, where $\eta_c = 0.1$ (unitless) and $\mu_0$ is the vacuum magnetic permeability (see \citeA{Jarvinen2018}). The inner boundary at which particles, planetary and solar, are removed is $100 \ km$ above the surface. We expect solar wind ions to have interacted with the planet's collisional atmosphere by this point. Table 1 shows simulation parameters and plasma parameters computed in the solar wind.

\begin{table}[ht!]
\color{black}
 \caption{A table of simulation values (global) and plasma parameters in the solar wind.}
 \centering
 \begin{tabular}{c c c c c c c c c}
 \hline
  Parameter & Symbol  & Value       \\
 \hline
   Timestep & $\Delta t$  & $ 0.026 \ s$     \\
   Grid resolution & $\Delta x$  & $ 411 \ km$     \\
   IMF vector & $B_{imf}$  & $ -5 \ nT \ \hat{z}$     \\
   Solar wind velocity & $\vec v _{sw}$  & $ -400 \ km/s \ \hat{x}$     \\
   Convection electric field & $\vec E _{sw}$  & $ 2 \ mV/m \ \hat{y}$     \\
   Solar wind temperature & $T_{sw}$  & $ 100,000 K$     \\
   Injected ion temperature & $T_{i}$  & $ 1,200 \ K$     \\
   Alfv\'en
 Mach number & $M_A$  & $ 8.2$     \\
   Sonic Mach number & $M_S$  & $ 10.8$     \\
  Magnetosonic Mach number &  $M_{ms}$  & $ 6.5$     \\
   Plasma Beta & $\beta$  & $ 0.69$     \\
   $H^+$ gyroradius & $r_{g,H^+}$  & $ 836 \ km$     \\
   $O^+$ gyroradius & $r_{g,O^+}$  & $ 13,265 \ km$     \\
   $H^+$ inertial length & $d_{i,H^+}$  & $ 101.7 \ km$     \\
   $O^+$ inertial length & $d_{i,O^+}$  & $ 406.8 \ km$     \\
 \hline
 \end{tabular}
 
 \end{table}

\hll{We set the emission radius to $R = R_E + 300 \ km$, which is above the peak ion densities at low and mid latitudes (see \citeA{Schunk2009}, Fig. 11.9), but below the exobase $500 \ km$ exobase \cite{Dowling2007}.} The dominant ion species at this altitude are $O^+$ and $H^+$ \cite{Glocer2009}. Planetary macro-ions are produced with velocity vectors normal to and upwards from the planetary surface. The temperature of injected ions at $300 \ km$ is pulled from a Maxwellian centered on $\sim 1,200 K$ (\citeA{Schunk2009}, Fig. 11.17), which corresponds to average thermal velocities of $1-3 \ km \ s^{-1}$ ($v_{th} = \sqrt{k_bTm^{-1}}$). \hll{The number of macro-ions per cell in the simulation are set to $50$, $10$, and $6$ for solar $H^+$, planetary $O^+$, and planetary $H^+$, respectively. The global rate of planetary ion emission} ($E_m$) is typically set by comparing model results to in-situ spacecraft measurements \cite{Kallio2002, Jarvinen2009}, however, such measurements do not exist for unmagnetized Earth. Thus, we explore a range of emission rates centered on reference rates for present-day Earth.

\subsection{Selection of Global Ion Emission Rates}
\hll{As seen in Eq 1, the dominant form of ion production should
be unaffected by demagnetization. We base our parameter space around a
reference set of emission rates $E_{ref}$, which are estimates based on present-day
measured outflows.} Measurements of the typical global $O^+$ outflow at Earth range from approximately $1-8\times 10^{25} \ s^{-1}$, but sometimes as high as $1\times10^{26} \ s^{-1}$ \cite{Nilsson2012,Yau1997, Lennartsson2004, Slapak2017}. However, during periods of low solar EUV flux, estimates of $O^+$ outflow at Earth have been reported as low as $2 \times 10^{24} \ s^{-1}$ \cite{Schillings2019PhD, Andersson2005}. Thus, we select the reference oxygen emission rate to be $2\times10^{25}\ s^{-1}$, which we vary by an order of magnitude in either direction across the simulations to include the range of measured rates.  

It is important to note that the cause and morphology of ionospheric outflows at various altitudes will look different between magnetized and unmagnetized planets \cite{Ramstad2021}. With the magnetized case, the outflows are concentrated at high latitudes \cite{Peterson2001} and are largely transported to the plasma mantle where they go on to escape (see \citeA{Slapak2017}, Figure 1). For unmagnetized planets, heavy ions escape from the day side in a plume in the convection electric field hemisphere, and there is a cold tail of escaping ions on the night side \cite{Fedorov2008, Dong2015}. Despite these differences, the outflow rates measured at Earth are comparable to those measured at Venus \cite{Ramstad2021}. At Venus, our nearest Earth analog, \citeA{Lundin2011} used \emph{Venus Express} data to derive a global $O^+$ escape rate of $1.2 \times 10^{25} \ s^{-1}$, which corresponds to an ion emission rate of roughly $8 \times 10^{25} \ s^{-1}$ in the RHybrid model (see Figure 10 of \citeA{Jarvinen2009}). While the differences between the two planet's ionospheres are significant, this comparison with Venus is additional evidence that we are modeling the proper range of $O^+$ emission rates at unmagnetized Earth in the present study ($2\times10^{24}\ s^{-1}$ to $2\times10^{26}\ s^{-1}$).

\hll{Escape of $H^+$ is usually estimated to be higher than that of $O^+$, for example see \citeA{Lundin2011} and Table 3 of \citeA{Peterson2001}. \citeA{Gunell2018} estimates that the outflow from an Earth-sized planet with today's intrinsic magnetic field to be $8\times10^{25} \ s^{-1}$ for $H^+$ and $2\times10^{25} \ s^{-1}$ for $O^+$ (see Fig. 3 therein).} Similarly, we set the reference emission rate of $H^+$ to be four times the emission rate of $O^+$ ($E_{ref} = [8\times10^{25} \ H^+s^{-1}, 2\times10^{25}\ O^+s^{-1}]$). We then run six simulations around these reference rates; each simulation is identical except for the input values of $E_m$. For $E_m$, we use the reference set multiplied by some factor $E_m =$ [$0.1E_{ref}$, $0.5E_{ref}$, $1E_{ref}$, $2E_{ref}$, $5E_{ref}$, and $10E_{ref}$]. \hll{The lower bound approaches the model's functional lower limit (see section 4.2.2 Net Mass Flux) and the upper bound is chosen such that the emission rate is log-symmetric about the reference values while capturing typical outflow rates measured at Earth and Venus.}

\subsection{Calculating the Escape and Deposition Rates}

We quantify global planetary ion escape rates in the simulation runs by subtracting the ion emission rate from the ion impact rate to the model's inner boundary for each individual species. The impact rate is the number of particles removed from the inner boundary every second; this value is tracked for planetary $H^+$, $O^+$ and solar $H^+$ separately. When the simulation reaches an effective steady state, planetary ions that are produced but do not return to the planet are assumed to have escaped.

The deposition rate of solar $H^+$ is simply the impact rate of solar $H^+$ onto the inner boundary at $100 \ km$ altitude. This is well into the planet's collisional atmosphere, so we assume that solar ions that reach this boundary have thermalized. The thermal velocity at this altitude is much lower than the escape velocity, therefore, deposited solar ions can be considered to be gravitationally bound. Some number of solar wind ions will have back-scattered, but it's possible that the majority are deposited \cite{Jolitz2017}. Depositing ions can also charge exchange multiple times with atmospheric particles as they penetrate \cite{Halekas2015}. In this study, they are simply removed at the inner boundary and considered to be deposited. Exospheric ion species are not included since the dominate planetary ion source near Earth is the ionosphere. There are no minor solar ion species included in the model runs.

The simulations yield time series data, which show how escape and deposition vary as a function of simulation time. Plots of these time series are shown in Appendix A. All simulation time series reach equilibrium. However, numerical and physical oscillations are still present. We find that the oscillatory behavior is well described by a normal distribution. We report the mean and standard deviation of the equilibrated time series for escape and deposition in Table 2.

\section{Results}

\subsection{Morphology of the Induced Magnetosphere}

\begin{figure}[ht!]
\centering
\makebox[\textwidth][c]{\includegraphics[width=1\textwidth]{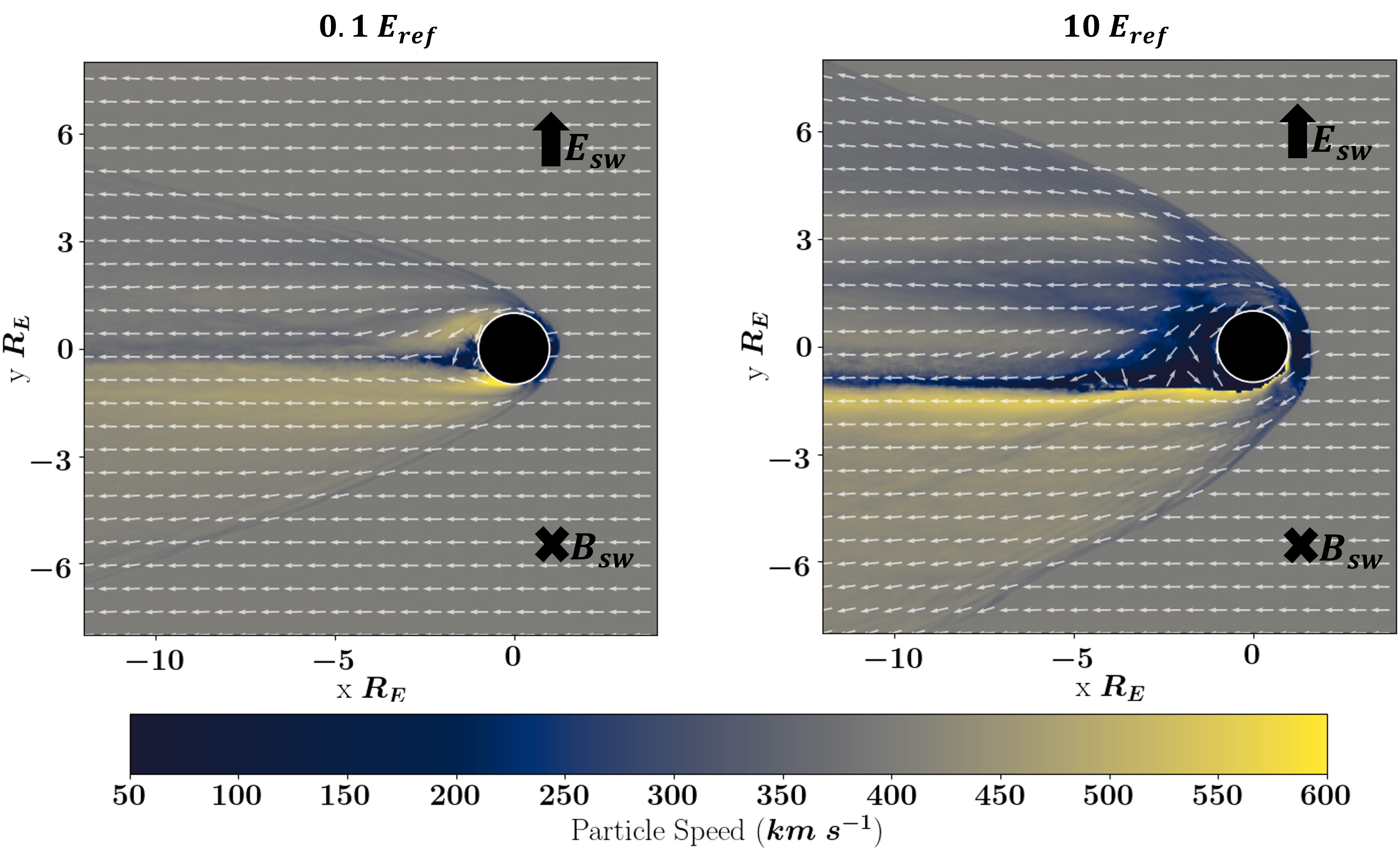}}
\caption{Snapshots in the xy plane from the $0.1E_{ref}$ (left) and $10E_{ref}$ (right) simulation runs. Each snapshot is taken from the end of the run when steady-state has been achieved. The blue circle at x =0, y =0, represents the planetary body. White overlaid vectors depict the direction of velocity vectors. The colorbar shows the speed of the solar wind, which is initialized at 400 km/s on the right side of the figures. The $10E_{ref}$ case has an inflated magnetosphere when compared to the $0.1E_{ref}$}
\label{Figure1}
\end{figure}

In all simulations, a magnetosphere with similar shape is induced around the planet. However, clear trends arise as the global ion emission rate is increased. Figure 1 depicts the solar wind velocity for two of the simulation runs: $0.1E_{ref}$ and $10E_{ref}$. These are the two extremes of this study, and the remaining simulation $E_m$ values lie between these two examples. Figure 1 shows that the induced magnetosphere is much larger in the case of higher ion emission. As more particles are injected into the planetary ionosphere the ionospheric pressure increases and pushes the pressure balance between the solar wind and ionospheric particles further away from the planet. Thus, the solar wind mass-loading also occurs further from the planet effectively inflating the magnetosphere \cite{Szego2000}. This causes the altitude of the bow shock at the planet-sun line to double from $0.3R_E$ to $0.6R_E$ in the $0.1E_{ref}$ and $10E_{ref}$ simulations, as shown in Figure 1. In both magnetospheres, the solar wind is shocked and decelerated in the subsolar region down to an average velocity of approximately $50 \ km/s$ in the magnetosheath.  Figure B1 in the Appendix shows the solar wind density, the flux of planetary ions, and the magnetic field for context.

The solar wind behavior (density and velocity) inside the magnetosphere is very similar across the different simulation runs. Directly behind the planet, in the xy plane, there is a slow solar wind ($10 - 70 \ km/s$) with a density of 0.1 to 2 protons $cm^{-3}$. In the $0.1E_{ref}$ case, small regions of acceleration exist near the planet on both the positive and negative y sides. However, the solar wind density in these regions is the lowest, between 0 and 1 protons $cm^{-3}$, so these acceleration regions do not host many particles by comparison. \hll{Similarly, in the $10E_{ref}$ case, a significant velocity enhancement to  $600 \ km/s$ occurs on the $-y$ side of the planet opposite the convection electric field. The particle density is very low in this region, so it does not represent a large part of the penetrating solar wind population.}

In both cases ($0.1E_{ref}$ and $10E_{ref}$), the curvature of the magnetic field is large near the pole in the -$E_{sw}$ hemisphere. This curvature results in a strong $\vec{J} \times \vec{B}$ force that re-accelerates the solar wind \cite{Tanaka1993}, as seen in \citeA{Jarvinen2013}. In each simulation, the electric and magnetic field vectors are oriented such that their cross product has a component directing particles along the magnetosheath. The electric fields in the $xy$ plane along the magnetosheath range from $2-3 mV/m$ for both cases. The magnetic fields along the sheath are about 20\% higher in the greater ion emission case. This leads to a stronger $\vec{E} \times \vec{B}$ drift in the $10E_{ref}$ simulation that better directs the ions away from the planet and the interior magnetosphere. This effect, combined with the greater stand-off distance, leads to the induced magnetosphere better shielding the planet from solar wind deposition as the ion emission rate is increased (see Figure 2 and Table 2).

\begin{figure}[ht!]
\centering
\includegraphics[width=25pc]{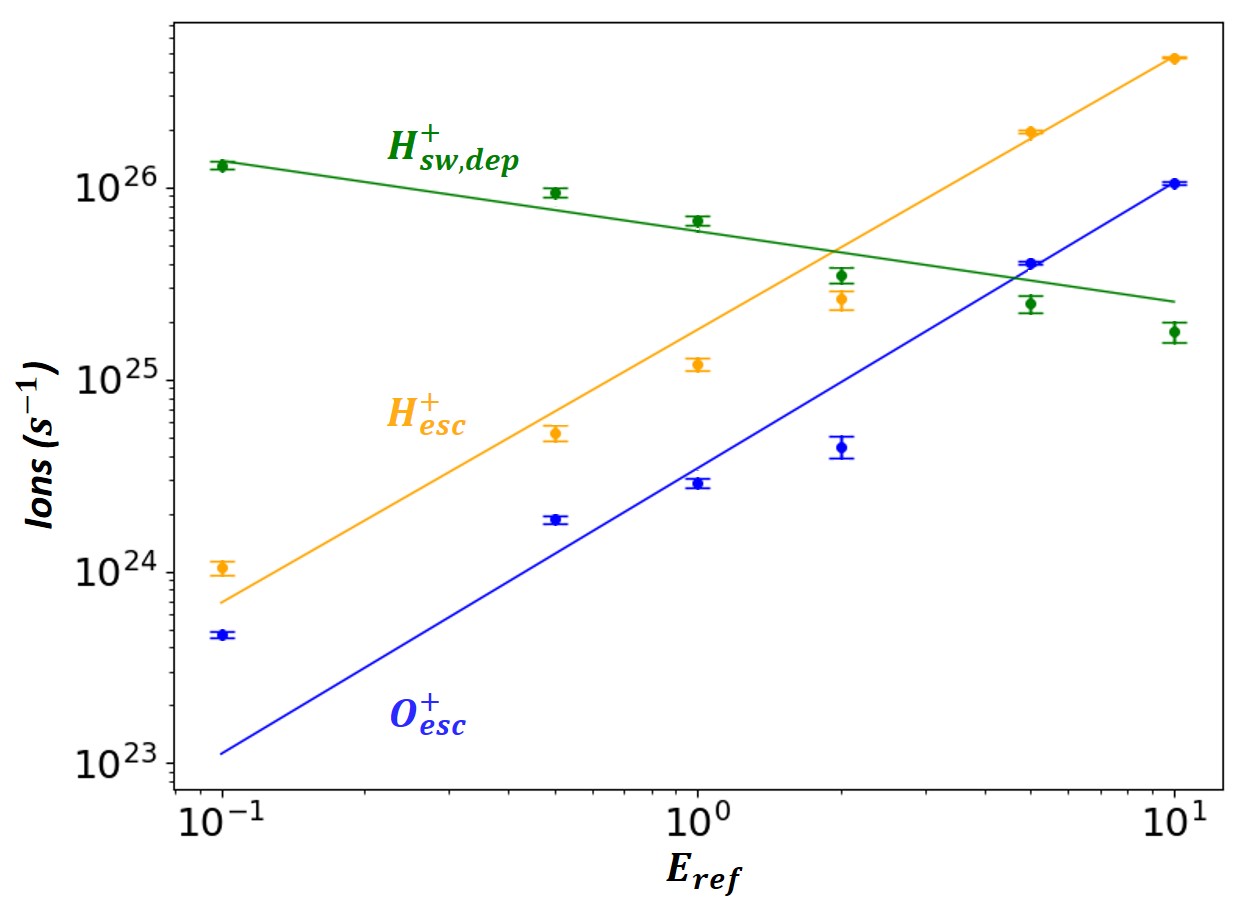}
\caption{The x-axis shows the global ion emission rate ($E_m$) in units of the reference rate ($E_{ref}$). The y-axis is in units of ions per second. The data with error bars is shown in log-log space, and overlaid with power laws. Orange and blue are hydrogen and oxygen escape, respectively. Green is solar hydrogen deposition. Error bars represent $1\sigma$ variation on Gaussian fits to equilibrated time series data.}
\label{Figure2}
\end{figure}

\subsection{Atmospheric Mass Flux}

\subsubsection{Absolute Escape and Deposition Rates}

The absolute number of escaping hydrogen ions and oxygen ions increases as the global ion emission rate increases, as shown in Figure 2. The increase roughly follows a power law distribution ($Q_m = aE_m^b$). The power laws for $H^+$ and $O^+$ shown in Figure 2 have constants $a = [1.81 \times 10^{25},3.43\times 10^{24}]$ and $b = [1.42, 1.49]$, respectively, where $E_m$ is in units of the reference rate for species $m$. The solar wind deposition rate decreases with increasing ion emission rate and is comparable in magnitude to the escape rates. This decrease also follows a power, which is plotted in Figure 2 as $a = 5.88\times 10^{25}$ and $b = -0.367$. At lower values of $E_{ref}$, the deposition rate can be greater than each of the individual escape rates. As the emission rate is increased, the number of escaping $H^+$ ions is equal to, and then exceeds, the number of deposited $H^+$ when $E \approx 2E_{ref}$. The number of escaping $O^+$ begins to exceed the number of deposited $H^+$ when $E \approx 4.6E_{ref}$. The escape and deposition rates, absolute and fractional, are listed in Table 2.

\subsubsection{Net Mass Flux}

In order to estimate the net mass flux of the planetary atmosphere, we calculate the total number of escaping $(Q)$ and depositing $(D)$ protons at each $E_m$ value. We assume that the total number of protons that escape (by mass) is the escaping hydrogen plus sixteen times the escaping oxygen, due to their respective atomic weights ($Q_{proton} = Q_{H^+}+16Q_{O^+}$). For each $E_m$ value, we subtract the total proton escape from the total proton deposition to compute the net atmospheric change, shown as the black triangles in Figure 3 ($\Delta N = D_{H^+} -Q_{H^+}-16Q_{O^+}$). In order to compute the error in these values, the errors for escape and deposition are added in quadrature. Explicit numerical values are listed in Table 2.

\begin{figure}[h]
\makebox[\textwidth][c]{\includegraphics[width=1.2\textwidth]{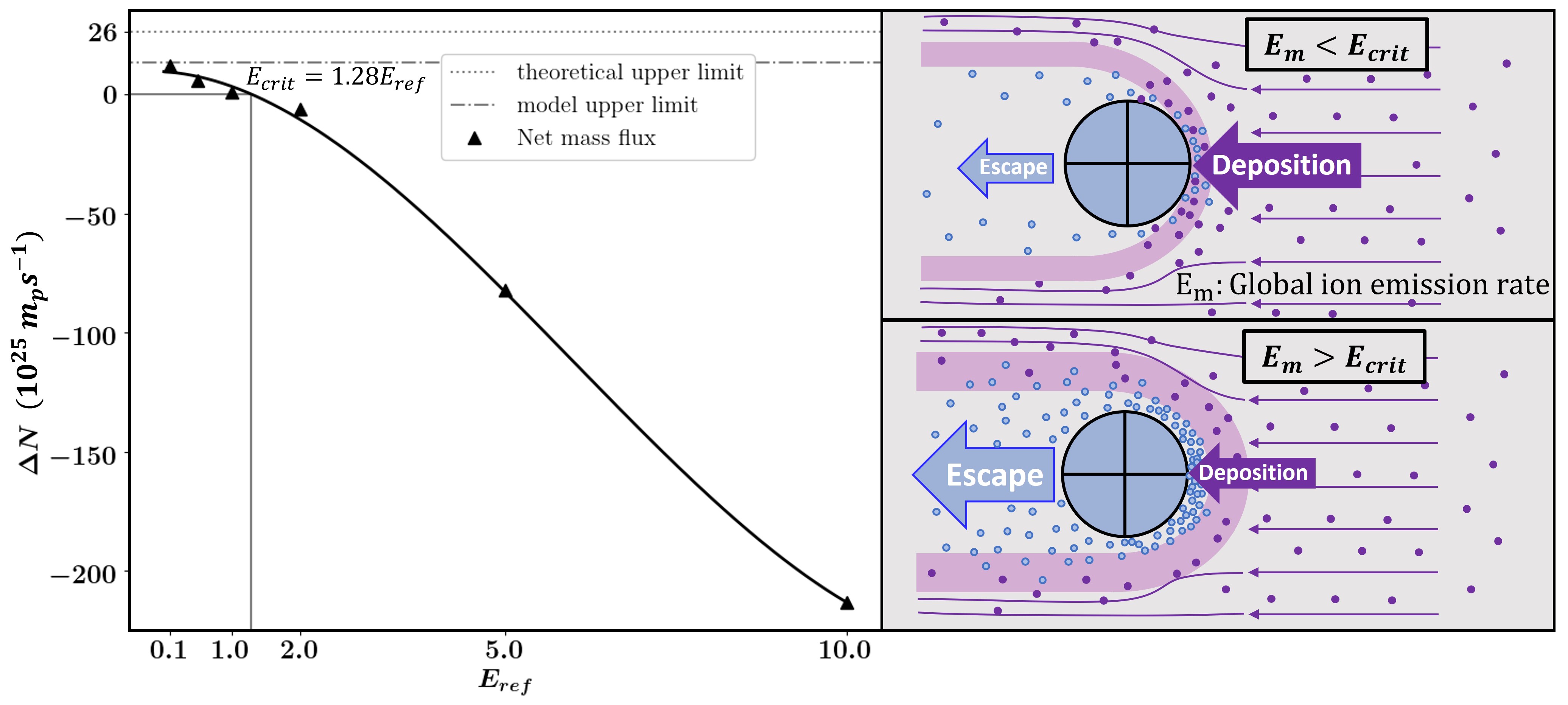}}%
\centering
\caption{In the left panel, the x-axis shows the global ion emission rate ($E_m$) in units of the reference rate ($E_{ref}$), and the y-axis shows the net mass change per second. A positive net mass change means that more mass is being deposited than is escaping, quantified in units of proton masses per time. The theoretical maximum limit on solar wind deposition is shown as the black dotted line and the model upper limit is shown as the black dash-dot line. Triangles represent computed net mass flux, and the thick black line is a polynomial fit to the data. Error bars are smaller than the marker size. The solid horizontal line represents a net mass flux of zero, and the solid vertical line represents the corresponding critical global ion emission rate. The right panel visualizes when $E_m$ is less than and greater than $E_{crit}$. Purple dots represent solar wind ions, blue dots represent planetary ions, and the light pink region is the magnetosheath.}
\label{Figure3}
\end{figure}

In Figure 3, the top dotted line is the direct flux of undisturbed solar wind $H^+$ ions on a geometric disk with $r = R_E$. This value can be taken as an upper limit, ignoring sputtering, back-scatter, and ion focusing and re-circulation towards the planet. This upper limit can be calculated for an arbitrary planet and stellar wind with $N_{max} = n_{sw}v_{sw} (\pi R_p^2)$, where $n_{sw}$ is the wind density ($5 \  cm^{-3}$), $v_{sw}$ is the wind speed ($400 \ km/s$), and $R_p$ is the radius of the planet ($R_E = 6,371 \ km$). For this study, $N_{max} = 2.6 \times 10^{26} \ protons/s$.

The horizontal dash-dot black line is the solar wind deposition rate without any planetary ion emission in a control run of the model. In this control, the model inner boundary is a superconducting sphere and ionospheric ion emission is switched off. The interplanetary magnetic field (IMF) drapes purely around the superconducting sphere and produces an induced magnetosphere, without mass-loading, that works to redirect some of the solar wind and causes the model limit to be lower than the theoretical limit.

As $E_m$ increases, the net mass flux changes from being positive to negative, crossing a critical $E_m$ value ($E_{crit}=1.28E_{ref}$) where the net mass flux of the planet is equal to zero. At this value, the incoming mass from solar wind deposition perfectly equals the outgoing mass from ion escape. As $E_m$ increases further from this value, the magnetosphere continues to inflate. Less solar wind is able to penetrate inward, whereas the number of escaping pick-up ions increases. This causes the net mass flux curve to turn sharply downwards. This result suggests that it's possible for unmagnetized planets to accrete net mass from the solar wind provided that neutral escape is sufficiently low, but the domain where net mass is lost is much larger.

All else being equal, our results imply that a planet's atmospheric mass evolution, considering only the ion flux, will depend on where it currently lies on its net mass flux profile. If $E > E_{crit}$, it will lose mass with time. If $E < E_{crit}$, it will gain mass. If $E = E_{crit}$, then the atmosphere of the planet will be in perfect equilibrium with the solar wind with regard to mass exchange. In this state, the mass source terms are equal to the loss terms. 

The current neutral escape rate of Earth is estimated to be $6 \times 10^{26} s^{-1}$ \cite{Gronoff2020}, which is higher than even the largest deposition rate found in this study ($1.3 \times 10^{26} s^{-1}$). Thus, while we find that net mass can be accreted on unmagnetized earth via electromagnetic interaction with the solar wind, thermal escape of neutrals is still large enough to offset the gains from ion deposition. However, our findings suggest that it is possible for a planet with a lower thermal escape rate to accrete more mass from the solar wind than it loses through combined ion and neutral escape if the global emission rate is low enough. \hll{Venus is one example where a cooler atmosphere leads to a relatively low thermal escape of $Q_{neutral} = 10^{25} s^{-1}$ \cite{Bertaux2007, Gunell2018}.}

\subsubsection{Extrapolating to Geologic Timescales}

To understand the implications of these results on the mass content of the atmosphere, we find the percent change in mass of the atmosphere per billion years for each of the simulation runs. We divide the total mass of the Earth's atmosphere ($5.1 \times 10^{18} \ kg$, \citeA{TrenBerth1994}) by the mass of a proton ($1.7 \times 10^{-27} \ kg$), in order to find the effective proton content of the atmosphere ($3 \times 10^{45} \ protons$). We then multiply our net flux rates by $10^9 \ yrs$, and calculate the percent change in total proton content of the atmosphere over a billion years. We find the proton content, i.e. effective mass, to change by only 0.13\%, 0.062\%, 0.01\%, -0.07\%, -0.86\% and -2.2\% for the $E_m$ values tested in increasing order (see Table 2).

\begin{table}[ht!]
 \caption{Number of escaping planetary hydrogen ($Q_{H^+}$) and oxygen ($Q_{O^+}$) ions, deposition of solar hydrogen ($D_{H^+}$), net proton flux ($\Delta N = D_{H^+} - 16Q_{O^+} - Q_{H^+}$), the percentage of injected hydrogen and oxygen that escapes (\ $Q_{m}/E_{m}$), the percentage of solar wind from an Earth-sized cross section that deposits into the atmosphere ($D_{H^+}/N_{max}$), and net atmospheric flux ($\Delta M$) with$^{(a)}$ and without$^{(b)}$ deposition. Uncertainties represent $1\sigma$ variation on Gaussian fits to equilibrated time series data.}
 \centering
 \begin{tabular}{c c c c c c c}
 \hline
  $E_m  \ $  & $Q_{H^+}$ & $Q_{O^+}$  & $D_{H^+}$ & $\Delta N$  \\ $(ions \ s^{-1})$ & $\ $ ($10^{25}s^{-1}$) $\ $ & ($10^{25}s^{-1}$) & ($10^{25}s^{-1}$) & $\ \ \  $ ($10^{25} protons \ s^{-1}$) $\ \ \  $   \\
 \hline
   $0.1 E_{ref}$  & $ 0.105 \pm0.009$ & $0.047 \pm0.002$ & $13.0 \pm0.6$ & $12.2 \pm0.6$    \\
   $0.5 E_{ref}$  & $0.53 \pm0.05$ & $0.187 \pm0.009$ & $9.4 \pm0.5$ & $5.9 \pm0.5$  \\
   $1 E_{ref}$  & $1.2 \pm0.1$ & $0.29 \pm0.02$ & $6.7 \pm0.4$ & $0.9 \pm0.5$    \\
   $2 E_{ref}$  & $2.6 \pm0.3$ & $0.45 \pm0.06$ & $3.5 \pm0.3$ & $-6 \pm1$   \\
   $5 E_{ref}$  & $19.5 \pm0.4$ & $4.07 \pm0.07$ & $2.5 \pm0.3$ & $-82 \pm1$  \\
   $10 E_{ref}$  & $47.4 \pm0.6$ & $10.5 \pm0.1$ & $1.8 \pm0.2$ & $-213 \pm2$ \\
 \hline
 \end{tabular}
 \begin{tabular}{c c c c c c }
 \hline
  $E_m \ $   & \ \ $Q_{H^+}/E_{H^+}$ \ \ & \ \ $Q_{O^+}/E_{O^+}$ \ \ &\ \ $D_{H^+}/N_{max}$ \ \ & \ \ $\Delta M^a $ \ \ & \ \ $\Delta M^b$ \ \\ $(ions \ s^{-1})$ & & & & $(10^9y)^{-1}$ & $(10^9y)^{-1}$ \\
 \hline
   $0.1E_{ref}$  & $13.1\%$ &$23.4\%$ & $50.1\%$   &$ 0.13\%$ & $-0.009\%$   \\
   $0.5E_{ref}$  &$13.1\%$ &$18.7\%$ & $36.1\%$   &$ 0.062\%$ & $-0.037\%$   \\
   $1E_{ref}$  &$15.0\%$ &$14.5\%$ & $25.9\%$   &$ 0.009\%$ & $-0.061\%$   \\
   $2E_{ref}$  &$16.4\%$ &$11.1\%$ & $13.4\%$   &$ -0.0066\%$ & $-0.1\%$   \\
   $5E_{ref}$  &$48.7\%$ &$40.7\%$ & $9.5\%$   &$ -0.86\%$ & $-0.89\%$   \\
   $10E_{ref}$  &$59.3\%$ &$52.4\%$ & $6.8\%$   &$ -2.2\%$ & $-2.3\%$   \\
 \hline

 \end{tabular}
 \end{table}

While this study has focused on ions as mass source and loss terms, turning off Earth's intrinsic magnetic field would not immediately alter the thermal escape rate of hydrogen ($6 \times 10^{26} s^{-1}$, \citeA{Gronoff2020}), which is the dominant neutral escape mechanism. Therefore, via neutral escape, we expect unmagnetized Earth to lose an additional $0.64\%$ of atmospheric mass per billion years. This result suggests that the total mass content of unmagnetized Earth's atmosphere can be fairly stable, changing by $<3\%$ over one billion years via atmospheric neutral and ion escape under steady solar driving conditions. This general conclusion remains true even if the deposition rates are not considered.

Changes in chemical composition are also an important part of atmospheric evolution, namely because of potential impacts for habitability \cite{Terada2009}. These results show that even if atmospheric escape is in equilibrium with deposition, the atmosphere can be losing heavy ions (oxygen) and gaining light ions (hydrogen) resulting in changing relative volatile inventories. This implies a mechanism for shifting the redox state of a planetary atmosphere with time, and may be important for understanding whether conditions are appropriate for formation of prebiotic molecules \cite{Koyama2021}. However, deposition to the upper atmosphere may not translate to significant changes of the near-surface tropospheric composition; this will depend on the pathways of deposited solar ions. Furthermore, our extrapolation of the net mass flux rate over a billion years assumes that there will be no feedback as mass is lost or gained.

\section{Discussion}
\subsection{Feedback from Net Atmospheric Flux}

We show that if the mass flux rates from/to the upper atmosphere are constant on geologic timescales, the total mass of the atmosphere remains stable. However, it is unlikely that escape and deposition rates would remain constant. Changes in the solar EUV flux \cite{Cnossen2007} and changes in the temperature of the upper atmosphere (e.g. \citeA{Ma2023}, Fig. 2) can both strongly influence escape rates. It is also possible that a nonzero net mass flux itself can cause feedback which affects planetary ion production and escape.

\begin{figure}[h]
\centering
\includegraphics[width=25pc]{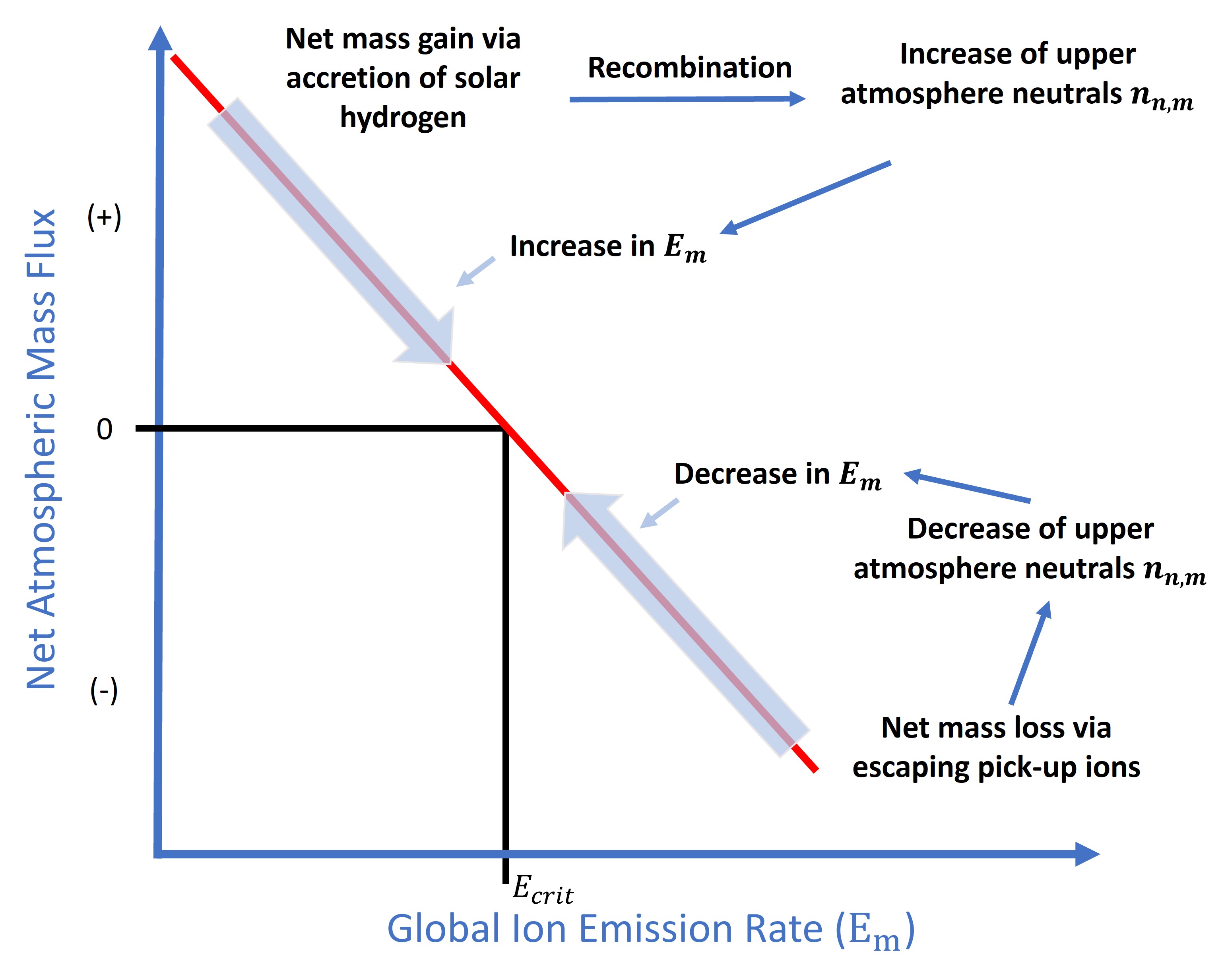}
\caption{This illustration shows how the modeled solar wind regulates $E_m$, and the total mass, of unmagnetized terrestrial atmospheres. Deviations from $E_{crit}$ cause negative feedback pushing $E_m$ towards $E_{crit}$.}
\label{Figure4}
\end{figure}

Consider a simplified scenario where the peak ion production along the planetary limb occurs at height $h_{0,1}$ above the surface according to a Chapman profile \cite{Chapman1931}. In the case of a positive upper atmospheric net mass flux, the atmospheric neutral density may increase over time, which can result in a greater column density of neutrals above $h_{0,1}$. These neutrals would absorb photoionizing flux, causing the peak ion production at $t_2 = t_1 + dt$ to shift in altitude to $h_{0,2}$, where $h_{0,2} > h_{0,1}$. At any given time the planetary atmosphere represents a photon-absorbing disk with an area that can be approximated by $A(t)=\pi (R_p+h_0(t))^2$. Expanding this equation and taking the derivative yields $dA(t)/dt=2\pi R_p(dh_{0}(t)/dt)$ (in the limit that $R_p >> h_0(t)$). Thus, the change in global photo-ionizing flux absorbed ($d\Phi/dt$) by a plane parallel atmosphere can be expressed as

\begin{linenomath*}
\begin{equation}
\frac{d\Phi}{dt}=2\pi R_p(dh_{0}(t)/dt)\int_{0}^{\lambda_M} F_\lambda \,d\lambda \hspace{10mm} [photons \ s^{-1}].
\end{equation}
\label{Eq2}
\end{linenomath*}

Equation (2) states that, all else being equal, the total number of ionizing photons absorbed varies in proportion with the change in altitude of peak ion production. Increased $h_0$ would lead to
a higher global ion emission, decreasing the net atmospheric mass flux.
Conversely, at high ion emission rates, the net mass flux of the planetary atmosphere is negative, meaning that more ions are escaping than being deposited. If this loss is not offset by a positive flow from lower atmospheric layers, then the neutral density in the upper atmosphere can decrease. Decreasing the upper atmosphere's neutral density can in turn decrease the cross-section of the photon-absorbing obstacle and the global emission rate $E_m$. This analysis suggests that deviations from $E_{crit}$ can have negative feedback, causing the interaction of an ionosphere with the solar wind to regulate global ion emission towards $E_{crit}$. A diagram outlining this feedback is illustrated in Figure 4. 

This negative feedback mechanism would only be operative on planets where ion escape and deposition are large compared neutral escape (e.g. cold upper atmospheres such as Venus). To further analyze this feedback mechanism, studies such as \citeA{Jolitz2017} can be carried out to determine at what altitude deposited ions of different energies will thermalize, and also what percentage of penetrating ions are thermalized versus back-scattered into the solar wind. From these results, the effect on the neutral density profiles can be obtained, and the magnitude of $dh_0(t)/dt$ can be reasonably estimated. It may turn out that this feedback mechanism has a much smaller effect on $E_m$ compared to other competing mechanisms on geologic timescales.

\subsection{Limitations}
This study considers solar $H^+$ deposition as a mass source term and pick-up ion escape as the loss term. Realistically, there is also solar $He^{++}$ deposition, thermal escape of neutrals \cite{Brain2016}, and the planet may be off-gassing additional mass into the atmosphere. The complete balance between sources and losses can be written as $D + S_{out} = Q_{ion} + Q_{neutral} + S_{absorb}$, where $D$ is the mass flux from solar deposition, $S_{out}$ is the combined out-gassing from surface sources that propagates to the ionosphere, $Q_{ion}$ is ion escape to space, $Q_{neutral}$ is neutral escape to space, and $S_{absorb}$ is the combined downward diffusion from the ionosphere. Here, we have effectively balanced this equation for the ionosphere via numerical modeling in the case where $S_{out} \approx S_{absorb}$ and where contributions from $Q_{neutral}$ are negligible.

Incorporating neutral escape would have the effect of decreasing the value of $E_{crit}$. For example, if $Q_{neutral} \approx 0.5Q_{H^+}$, we recompute $E_{crit}$ in this study to be $0.84E_{ref}$. However, both thermal escape of neutrals and the global ion emission rate depend on neutral density profiles \cite{Opik1963}. Properly including neutrals requires accurate coupling of the global ion emission rate with neutral densities. While including $Q_{neutral}$ would decrease $E_{crit}$, intrinsic planetary sources ($S_{out}$) would increase $E_{crit}$. Examples of intrinsic planetary sources that can propagate to the ionosphere are volcanism \cite{Themens2022, Yoshikawa2017} and sublimation of surface volatiles \cite{Hess1979}.

Another limitation of this study has to do with the selection of the hydrogen ion emission rates. In each simulation, we set the $H^+$ emission rate to be four times the $O^+$ emission rate. This is based on in-situ escape observations at Venus \cite{Lundin2011} and Earth \cite{Peterson2001}, and escape estimates for an unmagnetized Earth-like planet \cite{Gunell2018} (see Section 3.3 above). It is noted that oxygen ions can be more dominant than hydrogen ions at lower altitudes on Earth today \cite{Glocer2009}, thus, it is possible that we are overestimating the number of hydrogen ions emitted. \hll{However, as oxygen ions are the primary mass carriers in this study, it can be seen from Table 2 that removing the escape of hydrogen ions would not significantly affect the net mass flux balance.} 

This study provides a theoretical framework for understanding the mass flux and evolution of already established atmospheres on unmagnetized rocky planets. However, the dynamics of solar wind deposition have been simplified. It is thought that solar wind ions can charge exchange multiple times as they penetrate a planet's magnetosphere \cite{Halekas2015, Kallio1997}. As a result, neutral atmospheric particles can become ionized, and some will escape the planet as pick-up ions \cite{Brain2016}. Furthermore, some of the penetrating solar wind population will be scattered back \cite{Kallio2001}, while the rest may thermalize in the atmosphere \cite{Jolitz2017}. As they thermalize, they may stimulate auroral emission, such as the patchy proton aurora observed at Mars \cite{Chaffin2022} or the diffuse global aurora seen at Venus \cite{Gray2025}. To understand the detailed balance of escape and deposition on a specific planet, such as Mars or Venus, the details of these interactions must be rigorously incorporated. A review of the relevant processes can be found in \cite{Ma2008}. 

While we have included kinetic physics and a constant solar wind, future work involves adding additional processes (e.g. self-consistent ionospheric photochemistry, exospheric ion production, charge-exchange, and impact ionization) one at a time to understand the role of each in determining the plasma interaction at unmagnetized planets. In-situ data can be also analyzed to quantify solar wind populations that penetrate beyond the magnetopause at Mars and Venus. Future work will also explore how these results change as a function of solar wind velocity, density, and interplanetary magnetic field strength.

\subsection{Summary}
The role of intrinsic magnetic fields in atmospheric escape is multi-faceted and unresolved \cite{Brain2016b, Brain2023, Ramstad2021}. This study suggests that it is possible for an Earth-sized planet at $1 \ AU$ with no intrinsic magnetic field to experience modern solar driving conditions and maintain a stable atmosphere (by mass) for a billion years. All else being equal, the atmosphere and the solar wind may even evolve in the direction of mass equilibrium. Whether or not Earth's intrinsic magnetic field provides a shielding effect, these results indicate that the intrinsic magnetic field may not be necessary for atmospheric retention. Increasing temperatures due to increased solar intensity appear to pose a more significant threat to Earth's atmospheric retention than a cessation of the geodynamo (\citeA{Schoder2008} estimate the Earth's oceans will begin to boil off in about 1 billion years). Studies on how the atmosphere's composition would change may yet prove to be important to questions of habitability. Future work involves studying these compositional changes, as well as the upper atmospheric chemistry, in order to determine the possible pathways for deposited solar ions to diffuse downward into the troposphere.

The main takeaways of this study are as follows: [1] Precipitating ion fluxes may be comparable to ion escape fluxes and it is possible for the upper atmospheres of unmagnetized planets to accrete net positive mass from the solar wind. [2] Increased global ionospheric emission leads to an inflated induced magnetosphere, via solar wind mass-loading, that more efficiently deflects the solar wind around the planet. [3] For the solar wind conditions tested in this study, the mass content of unmagnetized Earth's atmosphere is stable within a few percent over a billion years. This remains true regardless of whether or not deposition is included in the calculation. [4] For unmagnetized planets in stellar winds, this study suggests that there exists a critical global ion emission rate $E_{crit}$ at which the upper atmosphere is in mass equilibrium with the solar wind (considering only ion flux). [5] A feedback loop is proposed that suggests a positive/negative net mass flux can inflate/deflate the upper atmosphere, which may regulate ion production towards a critical value where the net mass flux is equal to zero.

\appendix
\newpage

\section{Time Series Plots}

\begin{figure}[ht!]
\centering
\includegraphics[width=25pc]{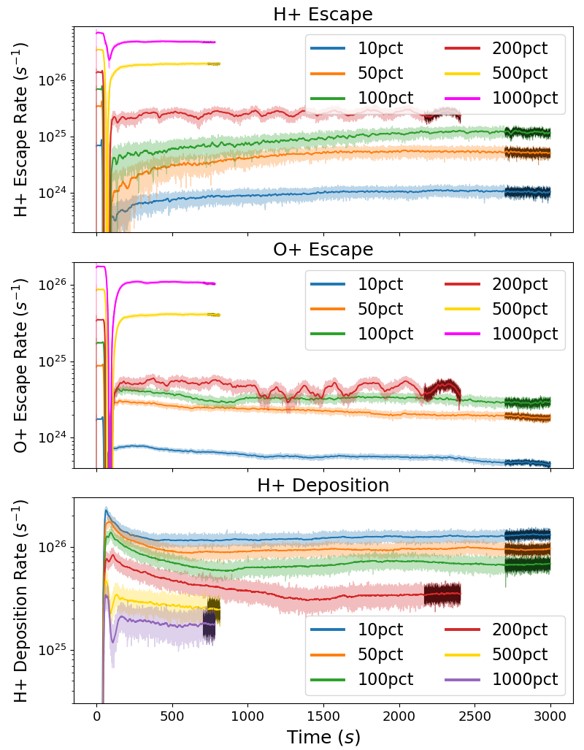}
\caption{The escape and deposition rates as a function of time are shown for each of the simulations. The top panel shows $H^+$ escape rates, the middle panel shows $O^+$ escape rates, and the bottom panel shows $H^+$ deposition rates. Higher production rate simulations require more computational time per real time step but reach equilibrium in less real time. The last 10\% of each time series (shown in black) is used to produce the data points in Figure 2. While the $2E_{ref}$ escape rate data is oscillatory, the sampled data captures an entire period, thus the effect of the oscillation is represented in the error bar of this data point.} 
\label{Figure5}
\end{figure}

\newpage

\section{Additional Model Visualizations}

\begin{figure}[ht!]
\centering
\includegraphics[width=25pc]{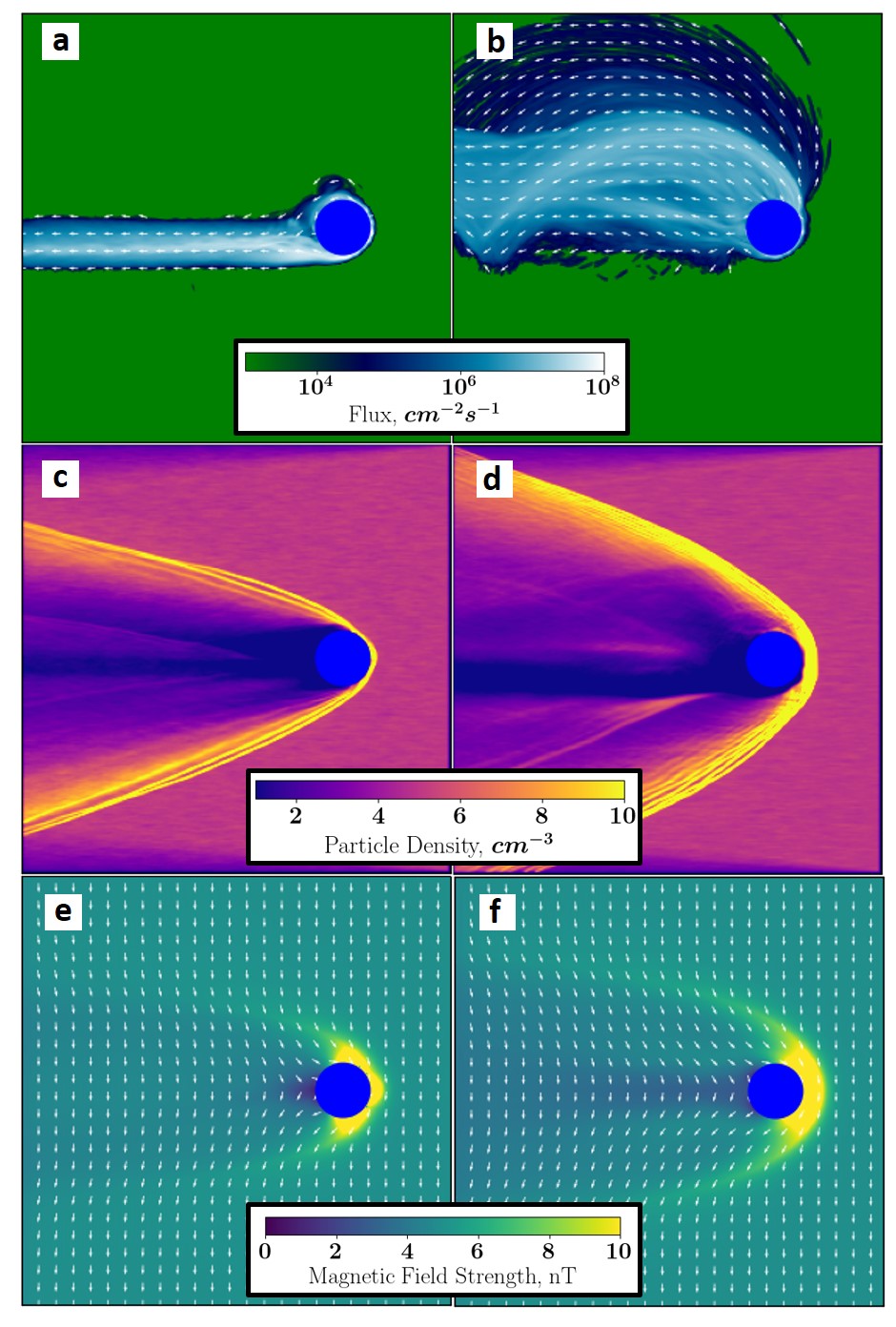}
\caption{Panels a and b show $H^+$ and $O^+$ flux, respectively, in the xy plane for the case where $E_m = E_{ref}$. Panels c and d show the solar wind density in the xy plane for $E_m = 0.1E_{ref}$ and $E_m = 10E_{ref}$, respectively. Panels e and f show the magnetic field in the xz plane for $E_m = 0.1E_{ref}$ and $E_m = 10E_{ref}$, respectively.} 
\label{Figure6}
\end{figure}

%
%
%
%

%



%
%

\section*{Open Research Section}
The data used in this study can be found by referring to Table 2. The model that generated these data, RHybrid, is open source and can be found at \url{https://github.com/fmihpc/rhybrid}, and is run with simulation parameters described in Section 2.3 of the present study. The simulation code version used in this study is archived \cite{jarvinen2024_rhybrid}. Global simulation results for each of the runs have also been archived \cite{hinton2025_simulationlog}.

\acknowledgments
This work utilized the Alpine high-performance computing resource at the University of Colorado Boulder. Alpine is jointly funded by the University of Colorado Boulder, the University of Colorado Anschutz, Colorado State University, and the National Science Foundation (award 2201538). Simulations were performed using the RHybrid code distributed under
the open source GPL v3 license by the Finnish Meteorological Institute
(github.com/fmihpc/rhybrid). Author Hinton thanks Andy Monaghan for his efforts in assisting with installing and running RHybrid on CU's super-computing clusters. Funding for this work was provided by the National Solar Observatory George Ellery Hale Fellowship and by the Retention of Habitable Atmospheres in Planetary Systems grant 80NSSC23K1358 awarded by the National Aeronautics and Space Administration’s Interdisciplinary Consortia for Astrobiology Research programme. RJ received funding from the European Research Council (Grant agreement No. 101124960).

%
%

\bibliography{bib.bib}

\end{document}